\documentclass[journal]{IEEEtran}
\usepackage[table]{xcolor}
\usepackage{pifont}
\usepackage{amsmath}
\usepackage{multirow}
\usepackage{textcomp}
\usepackage{booktabs}
\usepackage{subfigure}
\usepackage{amssymb}
\usepackage{graphics}
\usepackage{diagbox}
\usepackage{bm}
\usepackage{breqn}
\usepackage{pifont}
\usepackage{verbatim}

\newcommand{\oldmindcf}{$\mathrm{NDCF_{old}^{min}}$}
\newcommand{\newmindcf}{$\mathrm{NDCF_{new}^{min}}$}

\newcommand{\cmark}{\ding{51}}%
\newcommand{\xmark}{\ding{55}}%

\def\vec#1{\ensuremath{\bm{{#1}}}}

\ifCLASSINFOpdf
   \usepackage[pdftex]{graphicx}
   \graphicspath{{../pdf/}{../jpeg/}{resources}}
   \DeclareGraphicsExtensions{.pdf,.jpeg,.png}
\else
   \usepackage[dvips]{graphicx}
   \graphicspath{{../eps/}}
\fi

\begin{document}
%

\title{Speaker Recognition with Random Digit Strings Using Uncertainty Normalized HMM-based i-vectors}

%
%
%

\author{Nooshin~Maghsoodi, Hossein~Sameti,
        Hossein~Zeinali, and Themos~Stafylakis
\thanks{N. Maghsoodi and H. Sameti are with the Department
of Computer Engineering, Sharif University of Technology, Tehran 11365/8639, Iran (email: nmaghsoodi@ce.sharif.edu, sameti@sharif.edu).}
\thanks{H. Zeinali is with the Faculty of Information technology, Brno University of Technology, Brno, 61266, Czech Republic (e-mail: zeinali@fit.vutbr.cz).}
\thanks{T. Stafylakis is with the Computer Vision Laboratory, University of Nottingham, Nottingham, United Kingdom. (email: themos.stafylakis@nottingham.ac.uk).}}

\maketitle


\begin{abstract}
In this paper, we combine Hidden Markov Models (HMMs) with i-vector extractors to address the problem of text-dependent speaker recognition with random digit strings. We employ digit-specific HMMs to segment the utterances into digits, to perform frame alignment to HMM states and to extract Baum-Welch statistics. By making use of the natural partition of input features into digits, we train digit-specific i-vector extractors on top of each HMM and we extract well-localized i-vectors, each modelling merely the phonetic content corresponding to a single digit. We then examine ways to perform channel and uncertainty compensation, and we propose a novel method for using the uncertainty in the i-vector estimates. The experiments on RSR2015 part III show that the proposed method attains 1.52\% and 1.77\% Equal Error Rate (EER) for male and female respectively, outperforming state-of-the-art methods such as x-vectors, trained on vast amounts of data. Furthermore, these results are attained by a single system trained entirely on RSR2015, and by a simple score-normalized cosine distance. Moreover, we show that the omission of channel compensation yields only a minor degradation in performance, meaning that the system attains state-of-the-art results even without recordings from multiple handsets per speaker for training or enrolment. Similar conclusions are drawn from our experiments on the RedDots corpus, where the same method is evaluated on phrases. Finally, we report results with bottleneck features and show that further improvement is attained when fusing them with spectral features.     
\end{abstract}


%
\IEEEpeerreviewmaketitle

\section{Introduction}
%
%
%
%
\IEEEPARstart{D}{uring} the last several years, i-vectors~\cite{dehak2011front} have become the dominant approach to text-independent Speaker Verification (SV). In i-vector based systems, utterances of arbitrary duration are mapped onto a low-dimensional subspace modelling both speaker and channel variability, which is estimated in an unsupervised way. The back-end classifier that is usually employed is a Probabilistic Linear Discriminant Analysis (PLDA) model which performs a linear disentanglement of the two dominant types of variability and enables the evaluation of likelihood ratios~\cite{ioffe2006probabilistic}~\cite{senoussaoui2011well}. The i-vector/PLDA approach, when trained and evaluated on large text-independent datasets (such as those provided by NIST~\cite{doddington2000nist}) has shown a remarkable consistency over the years in attaining state-of-the-art performance. More recently, neural architectures (e.g. x-vectors \cite{snyder2018x}) have managed to outperform i-vectors in most text-independent SV benchmarks, by employing recent advances in deep learning and aggressive data augmentation \cite{snyder2019speaker}.

In parallel, the great potential of voice biometrics in commercial applications and forensics has increased the need for methods yielding state-of-the-art results with utterances of short duration. However, a straightforward application of the i-vector/PLDA model to short utterances has been proven to be an inadequate solution~\cite{stafylakis2013text}. When utterances become shorter, variations due to differences in phonetic content can no longer be averaged out, as happens with utterances of long duration (e.g. $>$1min). There have been several efforts to propagate the i-vector uncertainty to the PLDA model, but they were only partially successful and the results were inconsistent across datasets~\cite{kenny2013plda,stafylakis2013text,cumani2014use}. 

Due to the moderate performance of i-vectors in the particular setting, text-dependent SV started attracting much attention. Text-dependent SV reduces the phonetic variations of short utterances by constraining their vocabulary to either (a) a fixed phrase, (b) a set of predefined phrases, or (c) random sequences of words coming from a specific domain, such as digits. The first two approaches yield superior performance in general, due to the matched order of acoustic events between training and run-time utterances, which prevents random and hard to model co-articulation effects from appearing. On the other hand, when speakers utter a predefined pass-phrase, the system becomes vulnerable to spoofing attacks (e.g. replay attacks), which have become a major threat to speaker recognition systems~\cite{evans20172nd}. Text-prompted SV with random sequences of words from a specific domain is less vulnerable to replay attacks (yet not immune to attacks created by Text-To-Speech and Voice Conversion systems\footnote{In this case, creating a random sequence of words from a prerecorded audio is more difficult due to co-articulation effects of words on each other, but not impossible.}\cite{todisco2019asvspoof}) and it is employed as a means to perform liveness detection.

In this paper, we primarily work with RSR2015 part III, aiming at enhancing the i-vector paradigm in text-prompted speaker recognition~\cite{larcher2014text, larcher2012rsr2015}. One of our main motivations is to develop a method for utilizing the i-vector uncertainty tailored to text-dependent and text-prompted SV. We show that by introducing the concept of average uncertainty, a simple and effective linear digit-specific transform can be derived, which can compensate for the i-vector uncertainty without the computational burden in training and evaluation introduced by other uncertainty propagation methods \cite{stafylakis2013text,kenny2016uncertainty}. Building on top of our previous framework on text-dependent SV with fixed phrases (\cite{zeinali2016deep, zeinali2016trans}) and text-prompted case~\cite{zeinali2015telephony}, we use digit-specific HMMs and i-vector extractors and we report an extensive experimentation with respect the front-end features (including bottleneck features), channel compensation, uncertainty-aware transforms, and backend approaches. To the best of our knowledge, the results we report are the best published on the challenging RSR2015 part III and constitute a strong baseline for newer deep learning methods (e.g. \cite{zhu2018self,huang2018angular}).  

The rest of this paper is organized as follows. In Section~\ref{sec.RW}, we provide a detailed review of the proposed approaches to text-dependent SV and i-vector uncertainty modelling. In Section~\ref{sec.ubm_i-vec}, our digit-specific subsystems are explained. In Section~\ref{sec.uncertain} we discuss different methods for performing uncertainty-aware channel compensation. The description of the dataset, experimental setup and results are given in Sections~\ref{sec.expsetup} and~\ref{sec.results}. Finally, Section~\ref{sec.conclusion} we provide a brief conclusion of our work and directions for future work.

\section{Related Work}
\label{sec.RW}
In this section we present and discuss some of the recent approaches that are related to our work, with emphasis to those involving text-dependent speaker verification and uncertainty modelling. 
\subsection{Text-dependent speaker verification}

There are several interesting approaches to text-dependent SV that have been proposed over the last few years. In~\cite{aronowitz2012text}, the authors examine a pass-phrase based system which is evaluated on the (proprietary) Wells Fargo text-dependent dataset. NIST datasets were also used for UBM training to overcome the small development set constraint. In~\cite{novoselov2014text}, experiments are conducted on the same dataset and the authors propose the use of a separate Gaussian Mixture Model (GMM) mean supervector for each digit, adapted from a common UBM. Extracted supervectors undergo Nuisance Attribute Projection (NAP) and are passed to a Support Vector Machine (SVM) classifier to compute the scores. The authors show that their method outperforms the one proposed in~\cite{aronowitz2012text} on the same dataset. 

In~\cite{kenny2015vector, stafylakis2015jfa, stafylakis2015text}, the authors propose a Joint Factor Analysis (JFA) approach to address the problem of SV with random digit strings, using RSR2015 part III for training and testing. JFA is employed as a feature extractor, built on top of a tied-mixture model, i.e. an HMM with shared Gaussians and digit-specific sets of weights. The tied-mixture model serves for segmenting utterances into digits, as well as for collecting digit-specific Baum-Welch statistics for JFA modelling. JFA features are either local (i.e. one per digit) or global (i.e. a single per recording), and in the former case each local feature in the test utterance is scored against the corresponding ones in the enrolment utterances. In~\cite{stafylakis2015jfa}, JFA features are passed to a joint density back-end (alternative to PLDA), while in~\cite{kenny2015vector} the i-vector mechanics are used to incorporate the uncertainty in the back-end. Finally, in~\cite{kenny2016uncertainty}, the authors apply the same uncertainty-aware back-end to individual Gaussian mixture components, resulting in 20\% error rate reduction on RSR2015 part III. 

In the aforementioned approaches, a digit-independent or adapted UBM is employed, spanning the whole acoustic space. However, obtaining a robust estimate of a JFA speaker vector (i.e. $\vec{y}$-vector) using merely the tiny amount of information contained in a single digit (as happens with the local JFA features) proved to be very hard, making subspace methods to yield inferior results compared to supervector-size features (i.e. $\vec{z}$-vector). To address this data scarcity problem, a new scheme for using i-vector in text-prompted SV is introduced in~\cite{zeinali2015telephony}, where word-specific UBMs and i-vector extractors are employed. These UBMs and i-vector extractors are of small size (64-component and 175-dimensional, respectively) as they cover only the phonetic content of each individual word. Following a similar approach, in~\cite{zeinali2016deep, zeinali2016trans} it is shown that i-vectors, when extracted using phrase-specific UBMs and i-vector extractors yield superior performance compared to JFA front-end features. 

\subsection{Modelling i-vector uncertainty}
The use of i-vector uncertainty in the back-end may yield notable improvement in SV with short utterances and several methods for making use of it have been proposed. In \cite{kenny2013plda,stafylakis2013text} the authors introduce a modified version of PLDA for propagating the i-vector uncertainty to the PLDA model and they derive an EM algorithm for PLDA training using utterances of arbitrary durations. Similarly, the use of the i-vector uncertainty in PLDA is investigated in~\cite{cumani2014use}, taking into account only the uncertainty in the test utterances (i.e. assuming long training and enrollment utterances). The authors in \cite{lin2017fast,Lin+2016} speed-up the uncertainty propagation method by grouping i-vectors together based on their reliability and by finding a representative posterior covariance matrix for each group. In \cite{ribas2015uncertainty}, the authors incorporate the uncertainty associated with front-end features into the i-vector extraction framework. Finally, in \cite{saeidi2015accounting}, an extension of uncertainty decoding using simplified PLDA scoring and modified imputation is proposed. The authors also employ the uncertainty decoding technique in Linear Discriminant Analysis (LDA) in~\cite{saeidi2015uncertain}.
\section{Digit-specific HMMs, i-vector extractors and scoring}
\label{sec.ubm_i-vec}

In this section we present our HMM-based UBM which we use to extract Baum-Welch statistics and the method for using these statistics to train digit-dependent i-vector extractors. The scheme extends our previous method developed on Persian months dataset~\cite{zeinali2015telephony}. In~\cite{zeinali2015telephony} we proposed a simple but effective scheme based on separate i-vector extractor for each word with a common i-vector pipeline. In this paper, the above scheme is adapted to random digit strings and enhanced by exploring different methods for modelling uncertainty and combining it with channel compensation.  

Note that in RSR2015 part III, the sequence of digits in each utterance is assumed to be known and therefore can be used during training and evaluation~\cite{larcher2014text}, but our methods can be extended to settings where the digit sequences should be estimated by an ASR system.
\subsection{Digit-specific HMMs}

It is generally agreed that HMMs are a more natural solution to text-dependent SV than GMMs \cite{stafylakis2016speaker}. The partition of the Gaussians into HMM states permits us to capture the speaker characteristics over segments corresponding to phrases and words, rather than over merely spectral areas, as happens with a UBM. The HMM corresponding to digit $d$ is parametrized by a collection of $S_d$ state-specific GMM emission distributions $\left(w^{(s_{d},c_{s})}, \boldsymbol{\mu}^{(s_{d},c_{s})}, \boldsymbol{\Sigma}^{(s_{d},c_{s})}\right)$ and a transition probability matrix $\boldsymbol{A}_d$. We index HMM states by $s_{d}$ and Gaussian components of the GMM corresponding to HMM state $s_{d}$ by $c_{s}$. 

We initialize a collection of $D=10$ digit dependent HMMs, each having $S_d=8$ states and $C_{d,s}=8$ Gaussian components in each state $s$. We use the subscripts to indicate their dependence on the digit $d$ and state $s$, respectively, although in practice we use a fixed number of $S_d$ and $C_{d,s}$. The overall number of Gaussian components of the digit-specific HMM is $C_d=\sum_{s=1}^{S_d} C_{d,s} = 64$. HMM training and segmentation into digits is performed using Viterbi training (i.e. a single alignment of frames to HMM states is considered), by concatenating the corresponding digit-dependent HMMs and utilizing the ``left-to-right, no skips'' structure. Therefore, the concatenated HMM corresponding to an utterance with $d$ digits has $d\times S_d$ states and $d\times C_d$ overall Gaussian components, and it is constructed by concatenating the corresponding digit-dependent HMMs. 

Once the HMMs are trained, we jointly perform (a) segmentation of utterance into digits, and (b) segmentation of each frame sequence assigned to a digit to digit-specific HMM states $s_{d}$. We apply Viterbi-based forced alignment to assign frames to HMM states and hence we estimate hard assignment of frames to HMM states. Then, given the estimated alignments $\{\alpha^{d}_{t}\}_{t=1}^T$ to digit-specific HMM states $s_{d}$, frame posteriors corresponding to GMM components of the specific state are computed as follows,

\begin{equation}
\gamma_{t}^{(s_{d},c_{s})} = \delta_{s_{d},\alpha_{t}} \frac{w^{(s_{d},c_{s})}{\cal N}(\mathbf{o}_t|\boldsymbol{\mu}^{(s_{d},c_{s})}, \boldsymbol{\Sigma}^{(s_{d},c_{s})})}{\sum_{s_d}\sum_{c_{s}} w^{(s_{d},c_{s})}{\cal N}(\mathbf{o}_t|\boldsymbol{\mu}^{(s_{d},c_{s})}, \boldsymbol{\Sigma}^{(s_{d},c_{s})})}
\end{equation}
where $\delta_{\cdot,\cdot}$ is the Kronecker delta function, and ${\cal N}(\cdot|\cdot,\cdot)$ is the probability density function (PDF) of the multivariate normal distribution. Note that in $c_{s}$ the dependence on $d$ is kept implicit.

The frame posteriors, together with the corresponding Gaussian components are used to extract zero and first order centralized statistics, $\mathbf{N}_d=[N^{(d,1)}, \ldots,N^{(d,C_d)}]^T$ and
$\tilde{\mathbf{F}}_d=[{\tilde{{\mathbf{f}}}^{(d,1),T}}, \ldots,{\tilde{{\mathbf{f}}}^{(d,C_d),T}}]^T$, which are computed by the following equations
\begin{eqnarray}
\label{eq.zero_order_stats}
N^{(d,c)} & = & \sum_{t=1}^{L} \gamma_{t}^{(d,c)} \\
\label{eq.first_order_stats}
\tilde{\mathbf{f}}^{(d,c)} & = & \sum_{t=1}^{L} \gamma_{t}^{(d,c)} \left(\mathbf{o}_{t} - \boldsymbol{\mu}^{(d,c)} \right).
\end{eqnarray}
In the above equations, $L$ is the number of frames of the utterance, $c$ is the index of mixture component of the $d$ digit-specific mixture model, $\mathbf{o}_t$ is the frame at time $t$ and $\gamma_{t}^{(d,c)}$ is the posterior probability that the $t$ frame has been emitted by the $c$ component. Note that once the frame-posteriors $\gamma_{t}^{(d,c)}$ are calculated the HMM structure is no longer required for extracting Baum-Welch statistics. Therefore, $c = 1,2,\cdots, C_d$ is used for indexing components of the {\em flattened} HMM, i.e. a GMM corresponding to the concatenated state-specific GMMs, having overall $C_d=64$ Gaussian components and rescaled weights so that they sum up to 1. The flattened HMM plays the role of the UBM in text-independent speaker recognition.

\subsection{Digit Dependent i-vector Extractor}

Due to the use of digit-specific HMMs as UBMs for collecting Baum-Welch statistics, all the following structures should also be digit-specific. This includes i-vector extractors, transforms applied to i-vectors as well as trainable back-ends (e.g. PLDA).

The supervector $\mathbf{M}_d$ of an utterance associated with a digit $d$ is assumed to be generated from the following equation

\begin{equation}
	\mathbf{M}_d = \mathbf{m}_d + \mathbf{T}_d \mathbf{y}_d\,,
	\label{eq1}
\end{equation}
where $\mathbf{T}_d$ is a low rank matrix representing the subspace spanning the dominant variability in the supervector space, and $\mathbf{m}_d$ is the supervector corresponding to the digit-specific flattened HMM. Moreover, $\mathbf{y}_d$ is a latent variable with standard normal distribution as a prior. Given the Baum-Welch statistics of an utterance, the posterior distribution of $\mathbf{y}_d$ is normal with mean and covariance matrix estimated as follows
\begin{eqnarray}
	\label{eq2}
	\mathrm{cov}(\mathbf{y}_d) &=& {(\mathbf{I} + \mathbf{T}_d^t \mathbf{\Sigma}_d^{-1}\mathbf{N}_d \mathbf{T}_d)}^{-1}\,,\\ 
	\label{eq3}
	E[\mathbf{y}_d] &=& \mathrm{cov}(\mathbf{y}_d) \mathbf{T}_d^t \mathbf{\Sigma}_d^{-1} \tilde{\mathbf{F}}_d,
\end{eqnarray}
where $\mathbf{N}_d$ and $\tilde{\mathbf{F}}_d$ are zero and centralized first order statistics (using the means of the corresponding digit-specific HMM), and $\mathbf{\Sigma}_d$ is a block diagonal covariance matrix obtained from the corresponding digit-specific HMM.

\subsection{Digit-specific scoring}
After extracting digit-specific i-vectors and applying a set of transforms (to be discussed in Sect. \ref{sec.uncertain}), scoring is also implemented in a digit-specific fashion, i.e. 

\begin{equation}
S_{e,t} = \frac{1}{|\mathbf{D}_t|}\sum_{d \in \mathbf{D}_t} \mathrm{Sim}(\bar{\mathbf{y}}^{e}_{d},\mathbf{y}^{t}_{d}|{\cal P}_d)
\label{eq:aver_scoring}
\end{equation}
where superscripts $e$ and $t$ indicate enrollment and test respectively, $\mathbf{D}_t$ is the set of digits appearing in the test utterance of the trial ($|\mathbf{D}_t|=5$ in RSR2015 part III), and $\mathrm{Sim}(\cdot,\cdot| \cdot)$ is a similarity measure on the i-vector space (e.g. cosine similarity, PLDA-based log-likelihood ratio, a.o.), which is a function of parameters ${\cal P}_d$ (e.g. transforms for channel compensation, PLDA parameters) and can also include score normalization. Finally, we use $\bar{\mathbf{y}}^{e}_{d}$ to denote the averaged enrollment i-vectors of the digit $d$, since there might be more than one i-vectors of the same digit in the enrollment side (e.g. three in RSR2015 part III).   

This scoring rule is identical to the ``local'' approach, proposed in \cite{stafylakis2015text}. The rationale is to break-down the utterances into segments of limited phonetic content (e.g. words, digits) in order to suppress the phonetic variability between enrollment and test segments. A caveat is that certain segments of the enrollment utterances are not used in each trial, as the test utterance may not contain all the words appearing in the enrollment. In the case of RSR2015 part III, about 50\% of the enrolment number of frames is used in each trial, since $\mathbf{D}_t=5$. 

\subsection{Differences between the proposed method and tied-mixture models}

Apart from certain similarities between our method and the one in \cite{kenny2015vector,stafylakis2015jfa,stafylakis2015text}, the two methods are substantially different. Aside from differences in (a) subspace modelling (i-vectors vs. JFA features), (b) linear transforms applied to i- or y-vectors, and (c) back-ends (cosine distance vs joint-density models) there are differences in the way frames are assigned to Gaussian components. We propose digit-specific HMMs of $C_d=64$ Gaussian components each, without sharing them between digits or states, while in the tied-mixture approach, all $C=512$ Gaussian components are shared between digits, with the weights being the only digit-specific set of parameters. As a results, digit-specific i-vectors (or $y$-vectors \cite{stafylakis2015text}) using tied-mixture models are extracted over highly sparse Baum-Welch statistics, and are therefore characterized by high posterior uncertainty. Moreover, in the tied-mixture model approach the HMM structure is merely employed for segmenting utterances into digits, while we propose digit-specific HMMs to segment each digit into $S_d=8$ subword units. As a result, the Gaussian components are localized in the joint temporal and spectral domain, while in the tied-mixture approach they are merely localized in the spectral domain, via a standard UBM.

\begin{figure*}[!t]
  \centering
  {\includegraphics[width=\textwidth]{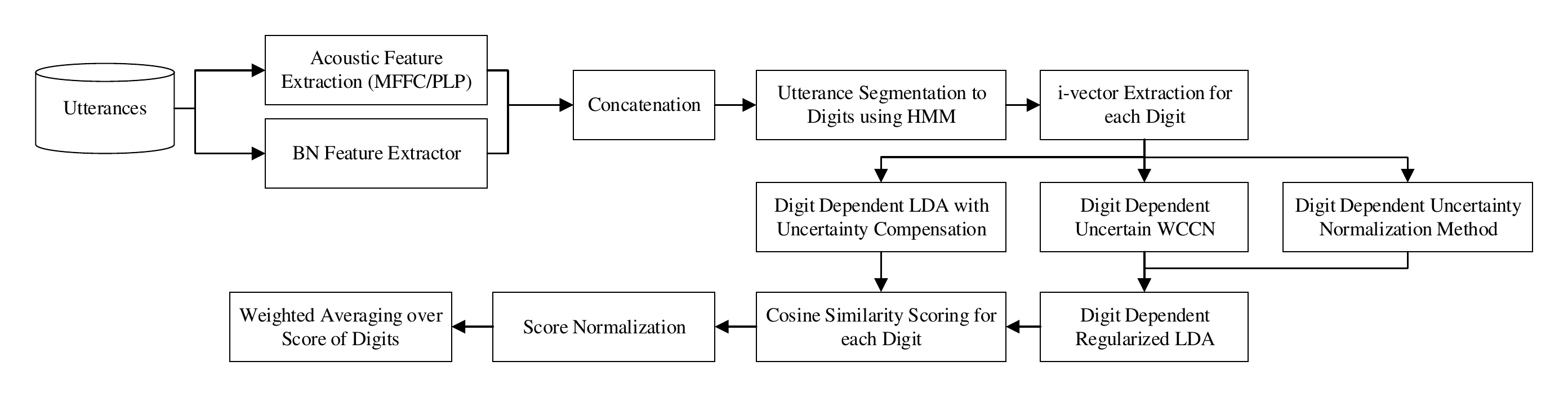}}
  \vspace{-6mm}
  \caption{\label{fig.fig1} Block diagram of the proposed system during enrollment phase.}
  \vspace{-3mm}
\end{figure*}


\section{i-vector uncertainty and channel compensation}
\label{sec.uncertain}

Due to the unsupervised way the i-vector extractors are trained, the i-vector space contains both speaker and session variability. Since only speaker information is useful to verify a speaker, a strategy for removing undesirable session effects is required. In parallel, in short duration SV the problem of increased uncertainty should also be addressed. To this end, we proposed three methods for channel and uncertainty compensation, which are explained in this section. \figurename~\ref{fig.fig1} illustrates the block diagram of the whole system, where all the examined compensation methods are depicted. In this figure, based on the selected method for uncertainty and channel compensation, one of the parallel switches is activated. 

\subsection{Between and within-class covariance and uncertainty}
It is well known that the total variability covariance matrix $\mathbf{S}_{tot}$ can be decomposed into between-class and within-class covariance matrices, $\mathbf{S}_b$ and $\mathbf{S}_w$ as follows
\begin{eqnarray}
    \label{eq4}
    \mathbf{S}_{tot} &=& \mathbf{S}_b + \mathbf{S}_w\,,\\
    \label{eq5}
    \mathbf{S}_b &=& \frac{1}{S} \sum_{s=1}^{S}(\overline{\mathbf{y}}_s - \overline{\mathbf{y}}) (\overline{\mathbf{y}}_s- \overline{\mathbf{y}})^T\,,\\
    \label{eq6}
    \mathbf{S}_w &=& \frac{1}{S} \sum_{s=1}^{S}{\frac{1}{n_s} \sum_{i=1}^{n_s}(\mathbf{y}_i^s- \overline{\mathbf{y}}_s) (\mathbf{y}_i^s- \overline{\mathbf{y}}_s)^T}\,.
\end{eqnarray}
However, by defining $\mathbf{S}_{tot}$ in the above way we are essentially treating i-vectors as point estimates. In order to take into account the uncertainty in the i-vector estimates, we should redefine the total variability as follows
\begin{dmath}
\label{eq_unc}
\mathbf{S}^{u}_{tot} = \frac{1}{n}\sum_{i=1}^{n} E\left[(\mathbf{y}_i - \overline{\mathbf{y}})(\mathbf{y}_i -\overline{\mathbf{y}})^T \right] = \frac{1}{n}\sum_{i=1}^{n} (E[\mathbf{y}_i] - \overline{\mathbf{y}})(E[\mathbf{y}_i] -\overline{\mathbf{y}})^T + \mathrm{cov}(\mathbf{y}_i) = \mathbf{S}_{tot} + \mathbf{S}_{u} ,
\end{dmath}
where $\mathbf{S}_{u} = \frac{1}{n}\sum_{i=1}^{n} \mathrm{cov}(\mathbf{y}_i)$ is the average uncertainty of the i-vectors and $n = \sum_{s=1}^{S}n_s$ is the overall number of i-vectors. The uncertainty in estimating $\overline{\mathbf{y}}$ is negligible as it is equal to $\frac{1}{n}\mathbf{S}_{u}$. It is interesting to note that $\mathbf{S}^{u}_{tot}$ is used in the i-vector extractor and in JFA during the minimum divergence estimation, where the latent variables are transformed in such a way so that $\mathbf{S}^{u}_{tot} = \mathbf{I}$. In other words, the covariance of the aggregated posterior $\mathbf{S}^{u}_{tot}$ is set equal to the covariance of the prior distribution by transforming $\mathbf{y}$ accordingly~\cite{dehak2011front}. The principal components of $\mathbf{S}_{u}$ correspond to the directions with the highest uncertainty. 

On the other hand, when dealing with short utterances, $\mathbf{S}_{u}$ becomes comparable to $\mathbf{S}_{tot}$ and it would be interesting to make use of it when performing channel compensation. We should moreover note that by decomposing $\mathbf{S}^{u}_{tot}$ into expected within and between-class covariance
\begin{equation}
\mathbf{S}^{u}_{tot} = \mathbf{S}^{u}_b + \mathbf{S}^{u}_w,
\end{equation}
we may consider $\mathbf{S}_{u}$ as being part of the within-class covariance, i.e.
\begin{equation}
\mathbf{S}^{u}_{w} \approx \mathbf{S}_w + \mathbf{S}_{u},
\end{equation} 
and
\begin{equation}
\mathbf{S}^{u}_{b} \approx \mathbf{S}_{b}.
\end{equation} 
This is due to the fact that the uncertainty contained in $\mathbf{S}^{u}_b$ is $\bar{n}$ smaller compared to $\mathbf{S}^{u}_w$, where $\bar{n} = \frac{n}{S}$ the average number of i-vectors per speaker, since
\begin{equation}   
E[(\overline{\mathbf{y}}_s - \overline{\mathbf{y}})(\overline{\mathbf{y}}_s - \overline{\mathbf{y}})^T] =  (\overline{\mathbf{y}}_s - \overline{\mathbf{y}})(\overline{\mathbf{y}}_s - \overline{\mathbf{y}})^T + \frac{1}{n_s}\mathbf{S}_s^{u},
\end{equation}
where $\mathbf{S}_s^{u}$ is the average uncertainty of i-vectors of speaker $s$.   

\subsection{Digit dependent Uncertainty and Channel Compensation}
We examine here our three different proposed approaches, as well as regularized LDA for applying session and uncertainty compensation. In all cases, the transformed vectors are obtains as $\mathbf{y}_i \leftarrow \mathbf{W}^T \mathbf{y}_i$.
\subsubsection{Uncertain LDA}
LDA is a standard technique to compensate for inter-session variability by finding a set of speaker-discriminant non-orthogonal directions and projecting the i-vectors onto the subspace they define~\cite{dehak2011front}. LDA minimizes the within-class variability while maximizing the between-class variability. Using the expectations of these matrices, the objective function of LDA becomes
\begin{equation}
	J(\mathbf{W})=\frac{\mathbf{W}^T \mathbf{S}^u_b \mathbf{W}}{\mathbf{W}^T \mathbf{S}^u_w \mathbf{W}} \approx \frac{\mathbf{W}^T \mathbf{S}_b \mathbf{W}}{\mathbf{W}^T (\mathbf{S}_w + \mathbf{S}_u) \mathbf{W}}. 
	\label{eq7}
\end{equation}
where $\mathbf{W}$ is the projection matrix.
By solving the above equation using generalized eigenvalue decomposition, uncertainty-aware channel compensation can be applied to i-vectors~\cite{saeidi2015uncertain}. 

\subsubsection{Digit dependent uncertain WCCN}

Within-Class Covariance Normalization (WCCN) is a popular technique for channel compensation that uses the Cholesky decomposition of the inverse within class covariance matrix \eqref{eq6} to project the input features. In speaker recognition, it is used typically before applying length normalization or cosine distance scoring~\cite{dehak2011front}. 
The uncertain version of WCCN is as follows
\begin{equation}
	(\mathbf{S}_w^u)^{-1} = \mathbf{W}\mathbf{W}^T,
	\label{eq14}
\end{equation}
where $\mathbf{W}$ is the projection matrix.
\subsubsection{Digit Dependent Uncertainty Normalization}
Finally, we propose a novel technique which we call Uncertainty Normalization. In this case, we are using only the average uncertainty and we ignore the clustering structure of i-vectors into speakers. It is an unsupervised method and therefore it does not require multiple recordings per speaker. The rationale is to project the i-vectors onto a space that down-scales directions exhibiting high uncertainty, since their estimates are less reliable. Similarly to uncertain WCCN, it is defined as follows 
\begin{equation}
	(\mathbf{S}^u)^{-1} = \mathbf{W}\mathbf{W}^T,
	\label{eq15}
\end{equation}
where $\mathbf{W}$ is the projection matrix.
\subsubsection{Regularized LDA} LDA has the constraint of reducing the dimensionality to at most $S-1$ where $S$ is the number of classes. Yet, in RSR2015 the number of training speakers is smaller than the i-vector dimension. To overcome this limitation and avoid dimensionality reduction we add a simple regularization term to $\mathbf{S}_b$. The regularized version of LDA yields better results than standard LDA in text-independent task too~\cite{zeinali2017sut}. In our experiments, we combine Regularized LDA with Uncertain WCCN and Uncertainty Normalization. 

\section{Experimental Setup}
\label{sec.expsetup}

\subsection{Datasets}

We used the RSR2015 part III dataset for almost all our experiments. In this dataset, there are 157 males and 143 females speakers, divided into three disjoint speaker subsets: background, development and evaluation, of about 100 speakers each. Each speaker model is enrolled with 3 10-digit utterances, recorded with the same handset, while each speaker contributes 3 different speaker models. Test utterances contain a quasi-random string of 5 digits, one out of 52 unique strings. Six commercial mobile devices were used for the recordings that took place under a typical office environment. All utterances are in English, while speakers are balanced in such a way so that they form a representative sample of the Singaporean population \cite{larcher2014text,stafylakis2015text}. 

Apart from the RSR2015 part III, two clean parts of 16\,kHz LibriSpeech dataset are used for training a DNN model and performing experiments with Bottleneck (BN) features (namely Train-Clean-100 and Train-Clean-360 \cite{panayotov2015librispeech}). The dataset contains English speech which is automatically aligned and segmented.

\subsection{Baseline and state-of-the-art}
As a baseline method, we refer to the experiments performed by CRIM (\cite{stafylakis2015text}) where both subspace and supervector domain methods are investigated. For fair comparison, we used the same setup as in \cite{kenny2015vector, stafylakis2015jfa, stafylakis2015text} and our baseline results are copied from the reference paper. The number of trials can be found in Table~\ref{tbl.trial_counts}.

To the best of our knowledge, the current state-of-the-art in RSR2015 part III is the model presented in \cite{zhong2017dnn}. The proposed system makes use of a DNN trained either on Fisher data or on RSR2015. Two main approaches are examined, namely DNN posteriors with MFCC features and tandem features, i.e. bottleneck features concatenated with MFCCs.    

\begin{table}[!t]
  \renewcommand{\arraystretch}{1.0}
  \caption{\label{tbl.trial_counts} Numbers of trials in RSR2015 part III based on the setup used in~\cite{kenny2015vector, stafylakis2015jfa, stafylakis2015text}.}
  \vspace{-2mm}
  \centerline
  {
    \begin{tabular}{l  c c}
      \toprule
      Set/Trial          & Male & Female \\
      \midrule
      Dev/target    	& 5134		& 4886 \\
      Dev/non-target	& 251381	& 224714 \\
      Eval/target	    & 5359  	& 416166 \\
      Eval/non-target	& 5188  	& 248852 \\
      \bottomrule
    \end{tabular}
  }
  \vspace{-5mm}
\end{table}

\subsection{Features}
We use 60-dimensional PLP or MFCC, extracted using HTK with a similar configuration: 25 ms Hamming windowed frames with 15 ms overlap. For each utterance, the features are normalized using Cepstral Mean and Variance Normalization (CMVN). A separate silence model is used for performing supervised Voice Activity Detection (VAD). Silent frames are removed after applying Viterbi alignment.

In addition to the cepstral features, a set of experiments is performed to examine the effectiveness of bottleneck (BN) and tandem features in the text-prompted task. To this end, a neural network is trained following the stacked architecture described and evaluated in~\cite{karafiat2014but,zeinali2017text}. Based on the reported results in~\cite{zeinali2017text}, this architecture exhibits very good performance in text-dependent SV. The output layer (softmax) has about 9000 senones, its input has 30 frames context around the current frame and it is trained using cross-entropy loss. Finally, the 80-dimensional BN features are concatenated to the cepstral features and used as input features to the i-vector pipeline.

\subsection{Model dimensions and gender dependence}
Digit-specific HMMs with 8 states and 8 components per state are used as UBM, while the i-vector dimensionality is set to 300. Gender independent UBMs and i-vector extractors are trained using only the background set of RSR2015. The background set is also used for training gender dependent LDA transforms as well as for score normalization. LDA and score normalization are applied in a digit-dependent manner. The MSR open source toolbox was used as a base for developing our code~\cite{sadjadi2013msr}. 

\subsection{Scoring method}

In our proposed system we use score-normalized cosine distance. As Eq. (\ref{eq:aver_scoring}) shows, for each digit-dependent test i-vector we extract, its cosine similarity with the average of the corresponding digit-dependent i-vectors from the enrolment speaker utterances is computed and the total score of the utterance is evaluated as the average score \cite{zeinali2015telephony}. It is worth mentioning that the proposed verification system uses a simple scoring method while other uncertainty-aware approaches typically require more complicated and computationally demanding methods, such as PLDA with uncertainty propagation~\cite{stafylakis2013text}. 

\subsection{Score and Length Normalization}

Score normalization is essential when cosine distance scoring is employed~\cite{dehak2011front}. After experimenting with several score normalization methods, we found that S-Norm yields the best performance. Therefore, for all the reported experiments and unless explicitly stated, S-Norm is applied in a gender and digit dependent manner, using the training set for collecting the cohort set of speakers. 

Although implicit in cosine distance scoring, length normalization helps towards obtaining more Gaussian-like distributions~\cite{garcia2011analysis}. It is therefore useful to apply it before LDA (and after uncertainty normalization), as the latter assumes Gaussian distributed class-means and class-conditional observations.

\section{Results}
\label{sec.results}

The evaluation metrics we report to assess the performance of the proposed methods are the Equal Error Rate (EER) and the Detection Cost Functions (DCFs) defined for NIST-SRE08 and NIST-SRE10, namely old Normalized DCF (\oldmindcf) and new Normalized DCF (\newmindcf).

\subsection{Baseline, state-of-the-art and our methods}

Table~\ref{tbl.results} shows the comparison between the proposed methods and several flavors of the baseline system. We select the best single system on this dataset from~\cite{stafylakis2015text} and fusion results of single systems with different combinations. \vec{y}-vector and \vec{z}-vector are JFA-features with and without speaker subspace, respectively. 

We also report results using speaker embeddings (x-vectors~\cite{snyder2018x}), which define the state-of-the-art in text-independent speaker recognition. The model attains state-of-the-art results on the Speakers In-The-Wild benchmark (namely 2.32\% EER on Eval Core \cite{mclaren2016speakers}). The x-vector architecture is trained using a large dataset with more than 7K speakers (VoxCeleb 1 and 2  \cite{nagrani2017voxceleb,chung2018voxceleb2}) compared to the 97 speakers used to train the i-vector extractors. All results reported are derived using identical evaluation set-up, ensuring a fair comparison.  

\begin{table*}[th]
  \renewcommand{\arraystretch}{0.95}
  \caption{\label{tbl.results}\it{Comparison of the proposed methods with baseline and state-of-the-art results.}}
  \vspace{-2mm}
  \centerline
  {
    \begin{tabular}{l l c c c c c c c c}
      \toprule
      & & & \multicolumn{3}{c}{\bfseries{Male}} & & \multicolumn{3}{c}{\bfseries{Female}} \\
      \cmidrule{4-6} \cmidrule{8-10}
      Model & Version & & EER [\%] & \oldmindcf & \newmindcf & & EER [\%] & \oldmindcf & \newmindcf \\
      \midrule
      \multirow{5}{*}{Baseline \cite{stafylakis2015text}}
            & Best \vec{y}-vector 		& & 3.63 & 0.194 & 0.632			&  & 5.66 & 0.288 & 0.762 \\
            & Fusion of \vec{y}-vectors 		& & 2.96 & 0.152 & 0.542			&  & 4.40 & 0.219 & 0.665 \\
        	& Best \vec{z}-vector  	& & 2.83 & 0.140 & 0.652			&  & 4.77 & 0.256 & 0.729 \\
        	& Fusion of \vec{z}-vectors 		& & 2.31 & 0.119 & 0.487			&  & 3.72 & 0.184 & 0.607 \\
        	& All Systems Fusion 	& & 2.01 & 0.105 & 0.491			&  & 3.19 & 0.162 & 0.553 \\
      \midrule
      \multirow{2}{*}{State-of-the-art \cite{zhong2017dnn}}
         & DNN i-vectors 	& & \bfseries{1.70} & 0.102 & -			&  & 2.69 & 0.150 & - \\
         & Tandem DNN i-vectors 	& & 1.81 & - & -			&  & \bfseries{1.84} & - & - \\
       \midrule
      \multirow{3}{*}{X-vectors}
            & PLDA trained on VoxCeleb		& & \bfseries{2.71} &  \bfseries{0.280} &  \bfseries{0.462}			&  &  \bfseries{2.88} &  {0.413} &  {0.710} \\
            & PLDA trained on RSR		& &{3.08} &  {0.342} &  {0.551}			&  &  {3.45} &  {0.415} &  {0.698} \\
            & PLDA adapted to RSR 		& & {3.03} &  {0.339} &  {0.553}			&  &  {3.31} &  \bfseries{0.411} &  \bfseries{0.693} \\
      \midrule
      \midrule
      \multirow{3}{*}{Proposed, MFCC}
          & Uncertain WCCN	      & & 1.76 & 0.103 & 0.504			&  & 2.12 & 0.106 & 0.458 \\          
          & Uncertain LDA      & & 1.84 & 0.106 & 0.532			&  & 2.19 & 0.112 & 0.465 \\
          & Uncertainty Normalization     & &  \bfseries{1.52} &  \bfseries{0.093} &  \bfseries{0.517}			&  &  \bfseries{1.77} &  \bfseries{0.094} &  \bfseries{0.424} \\
      \midrule
      \multirow{3}{*}{Proposed, PLP}
          & Uncertain WCCN	      & & 1.98 & 0.110 & 0.553			&  & 2.03 & 0.103 & 0.465 \\        	
        	 & Uncertain LDA      & & 2.20 & 0.124 & 0.585			&  & 2.41 & 0.126 & 0.520 \\
	 & Uncertainty Normalization     & & \bfseries{1.80} & \bfseries{0.103} & \bfseries{0.531}			&  & \bfseries{1.83} & \bfseries{0.094} & \bfseries{0.434} \\
      \midrule
      \multirow{3}{*}{Proposed, MFCC+PLP}
          & Uncertain WCCN	      & & 1.49 & 0.091 & 0.527			&  & 1.70 & 0.089 & 0.410 \\
        	 & Uncertain LDA      & & 1.64 & 0.097 & 0.534			&  & 1.95 & 0.096 & 0.433 \\
 	 & Uncertainty Normalization     & & \bfseries{1.45} & \bfseries{0.089} & \bfseries{0.528}			&  & \bfseries{1.65} & \bfseries{0.084} & \bfseries{0.395} \\        	
      \bottomrule
    \end{tabular}
  }
  \vspace{-4mm}
\end{table*}


\begin{figure}[tb]
  \centering
  \vspace{-6mm}
  {\includegraphics[width=7.5cm]{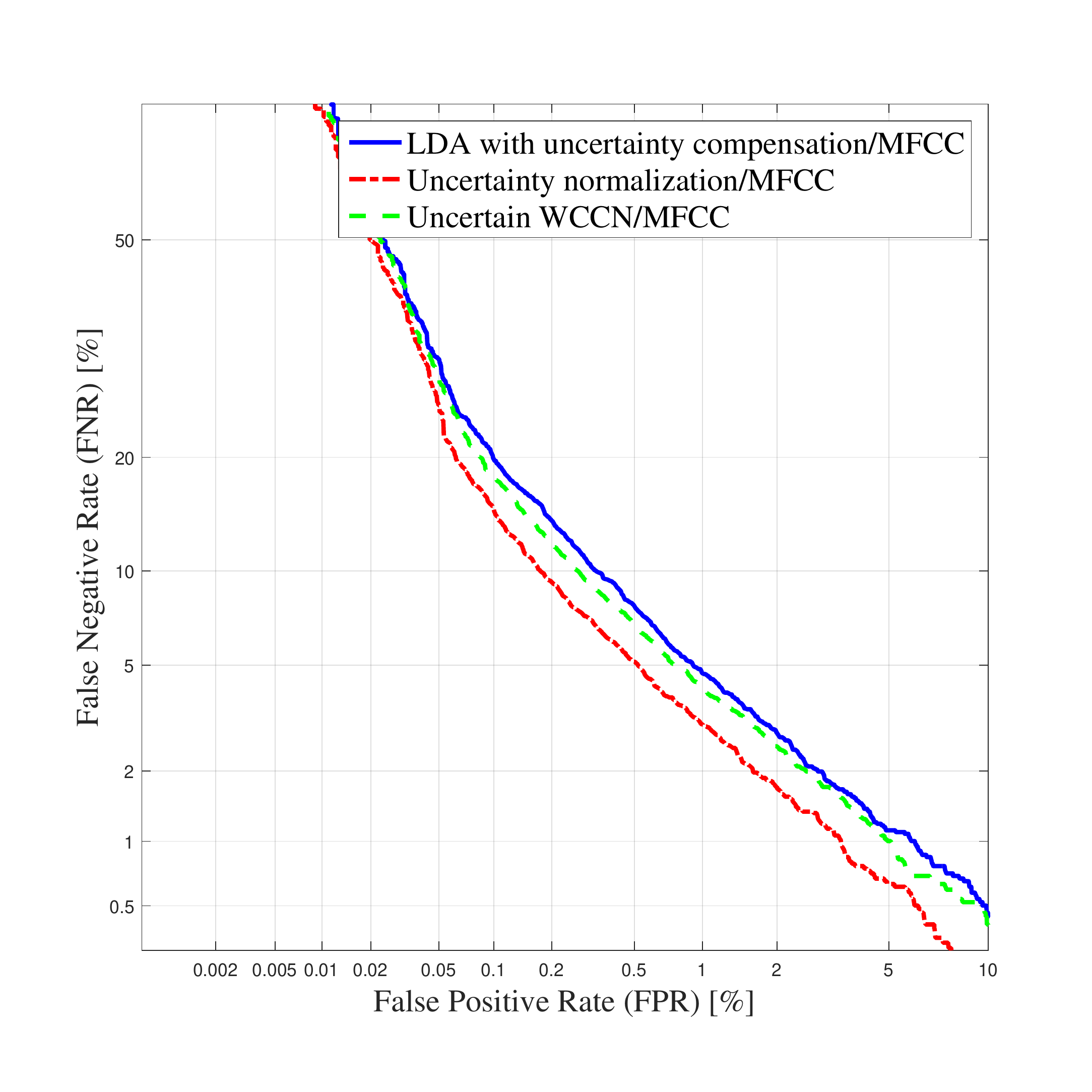}}
  \vspace{-8mm}
  \caption{\label{fig.det_female} DET curves for the proposed methods for female speakers. The trends for male speakers are similar.}
  \vspace{-4mm}
\end{figure}


In addition, \figurename~\ref{fig.det_female} shows the DET curves of some selected systems from Table~\ref{tbl.results} for female speakers. In the third and fourth sections of this table we report results for the systems with PLP and MFCC features. We observe that for both genders, MFCCs outperform PLPs in almost all experiments. Based on these results, the system with MFCC features is considered as the best single system. Moreover, score-level fusion results of the two systems are given the fifth section of Table~\ref{tbl.results}.

\subsection{Uncertainty Normalization, Channel Compensation and Score Normalization}

As Table \ref{tbl.results} shows, the proposed uncertainty normalization methods attain the best results. Hence, it is worth further analyzing its performance, e.g. by deactivating channel compensation (i.e. Regularized LDA) and score normalization. 

In Table \ref{tbl.resultsUN}, we report results using several such combinations, as well as an experiment with PLDA. First of all, we observe that the contribution of Regularized LDA is rather minor compared to uncertainty normalization. This result is rather surprising; it shows that state-of-the-art performance can be attained even without explicit channel modelling, i.e. without the need of collecting multiple training recording coming from different channels, sessions or handsets, per speaker. We mention again that in RSR2015 part III the enrolment utterances for a given speaker are coming from a single handset, which is different to the ones used in the test utterances \cite{larcher2014text}.

Finally, we examine the effectiveness of Gaussian PLDA as a backend. To this end, we train $D=10$ digit-dependent PLDA models using the RSR2015 part III training set. After experimentation, we found that the combination of uncertainty normalization, regularized LDA and number of speaker factors equal to 50 yields the best performance, while S-Norm does not yield any further gains. However, even the best PLDA configuration is clearly inferior to that attained by cosine distance. We believe that the failure of PLDA is due to the small number of training speakers in RSR2015, which prevents us from estimating robustly the speaker subspace. 

\begin{table*}[th]
  \renewcommand{\arraystretch}{1.0}
  \caption{\label{tbl.resultsUN}\it{Combinations of Uncertainty Normalization, Regularized LDA, S-Norm and PLDA}}
  \vspace{-2mm}
  \centerline
  {
    \begin{tabular}{l l c c c c c c c c}
      \toprule
      & & & \multicolumn{3}{c}{\bfseries{Male}} & & \multicolumn{3}{c}{\bfseries{Female}} \\
      \cmidrule{4-6} \cmidrule{8-10}
      Model & Version & & EER [\%] & \oldmindcf & \newmindcf & & EER [\%] & \oldmindcf & \newmindcf \\
      \midrule
      \multirow{5}{*}{Proposed}
            & Uncert. Norm, Reg. LDA, S-Norm 		& & \bfseries{1.52} &  \bfseries{0.093} &  \bfseries{0.517}			&  &  \bfseries{1.77} &  \bfseries{0.094} &  \bfseries{0.424} \\
            & Uncert. Norm, Reg. LDA		& & 2.04 & 0.113 & 0.533			&  & 2.57 & 0.133 & 0.515\\
            & S-Norm 		& & 2.15 & 0.113 & 0.546			&  & 3.12 & 0.143 & 0.561 \\ 
        	    & Uncert. Norm, S-Norm  	& & 1.68 & 0.093 & 0.550			&  & 1.89 & 0.102 & 0.440 \\
        	    & Uncert. Norm, Reg. LDA, PLDA  	& & 2.37 & 0.118 & 0.491			&  & 2.63 & 0.176 & 0.516 \\                                   	 
      \bottomrule
    \end{tabular}
  }
  \vspace{-4mm}
\end{table*}
\subsection{Comparison with x-vector}
The embedding extractor is implemented using the standard Kaldi recipe, and it is trained on VoxCeleb 1 and 2 (containing more that 7K speakers)~\cite{nagrani2017voxceleb}. The PLDA model used for evaluating LLRs is also trained on VoxCeleb, while we also report results where the RSR2015 training set is employed for PLDA training or adaptation. For enrolling the speakers, three utterances are concatenated and a single x-vector is extracted and for evaluation utterances, and each sequence is represented by an x-vector. 
The results in the third row of Table~\ref{tbl.results} show that the best performance is attained by training PLDA on VoxCeleb without any adaptation. However, our proposed method performs notably better than this. To improve the performance of x-vectors, recently proposed methods for applying domain adaptation to the x-vector extractor (e.g. using Generative Adversarial Networks \cite{rohdin2019speaker, nidadavolu2019cycle}) are worth exploring, in order to reduce the mismatch in channel and accent between VoxCeleb and RSR2015.

\subsection{Using bottleneck features}

Neural approaches using DNNs trained for ASR have resulted in significant improvements in SV, especially in text-independent SV, where the text is unknown and DNNs help towards assigning frames to ASR recognition units (e.g. senones)~\cite{lei2014novel}. Some recent works apply DNNs to text-dependent SV and report notable improvements~\cite{zeinali2016deep}~\cite{zeinali2017text}. Hence, it is worth examining the performance of bottleneck and tandem features extracted from a DNN in text-prompted case. Since the best performance among uncertainty and channel compensation methods is attained by uncertainty normalization followed by regularized LDA, we report the results using only this method. Table~\ref{tbl.bottleneck} shows the results obtained by 80-dimensional bottleneck feature vector, by their concatenation with MFCC feature vector (i.e. tandem features) and by their fusion with other cepstral features. The results show that although the performance of bottleneck features without any fusion is poor, fusing tandem features with other cepstral features yields significant improvements. The reason for this degradation could be the randomness in digit sequence compared to the fixed sequences of other text-dependent tasks, as well as the fact that we did not use in-domain RSR2015 data to fine-tune the network. It is also apparent from the results that tandem features yield more notable improvement for female speakers.

In Table \ref{tbl.stateNumber} we examine the performance of our best single-feature model (i.e. with MFCC) by varying the number of states per HMM $S_d$. As we observe, the performance is rather insensitive to $S_d$, being slightly higher for $S_d=16$. However, we choose to use $S_d=8$ for the rest of the experiments, since their differences are minor and the algorithm becomes less computationally and memory demanding.  
 
 \begin{table*}[th]
  \renewcommand{\arraystretch}{0.95}
  \caption{\label{tbl.bottleneck}\it{The results when using bottleneck features with uncertainty normalization.}}
  \vspace{-2mm}
  \centerline
  {
    \begin{tabular}{l l  c c c c c c c c}
      \toprule
      & & & \multicolumn{3}{c}{\bfseries{Male}} & & \multicolumn{3}{c}{\bfseries{Female}} \\
      \cmidrule{4-6} \cmidrule{8-10}
      Features & & & EER [\%] & \oldmindcf & \newmindcf & & EER [\%] & \oldmindcf & \newmindcf \\
      \midrule
      MFCC
          	   & & & 1.52 & 0.093 & 0.517			&  & 1.77 & 0.094 & 0.424 \\
      PLP
          	   & & & 1.80 & 0.103 & 0.531			&  & 1.83 & 0.094 & 0.434 \\
      \midrule
      BN
          	   & & & 3.80 & 0.210 & 0.666			&  & 3.48 & 0.173 & 0.578 \\
      BN+MFCC	
          	   & & &  2.99 & 0.158 & 0.582			&  & 1.97 & 0.098 & 0.405 \\
      BN+MFCC, MFCC
          	   & & & 1.51 & 0.091 & 0.506			&  & 1.33 & 0.066 & 0.360 \\
      BN+MFCC, PLP, MFCC
          	   & & & 1.38 & 0.086 & 0.987			&  & 1.23 & 0.062 & 0.349 \\
      BN+MFCC, BN, PLP, MFCC
               & & & 1.24 & 0.084 & 0.482			&  & 1.21 & 0.058 & 0.333 \\
      \bottomrule
    \end{tabular}
  }
  \vspace{-3mm}
\end{table*}

\begin{table*}[th]
  \renewcommand{\arraystretch}{1.0}
  \caption{\label{tbl.stateNumber} Comparison between different number of HMM states.}
  \vspace{-2mm}
  \centerline
  {
    \begin{tabular}{l l  c c c c c c c c}
      \toprule
      & & & \multicolumn{3}{c}{\bfseries{Male}} & & \multicolumn{3}{c}{\bfseries{Female}} \\
      \cmidrule{4-6} \cmidrule{8-10}
      Number of HMM states & & & EER [\%] & \oldmindcf & \newmindcf & & EER [\%] & \oldmindcf & \newmindcf \\
      \midrule
      \multirow{1}{*}{4}
            	& & &  1.59 &  0.096 &  0.519	&  &  1.82 & 0.097 & 0.431 \\
      \multirow{1}{*}{8}
                & & &  1.52 &  0.093 &  0.517	&  &  1.77 & 0.094 & 0.424 \\
      \multirow{1}{*}{16}
                & & &  1.50 & 0.089 & 0.516		&  & 1.76 & 0.089 & 0.422 \\
      \multirow{1}{*}{32}
                & & &  1.56 & 0.094 & 0.520		&  & 1.78 & 0.092 & 0.428 \\
      \bottomrule
    \end{tabular}
  }
  \vspace{-3mm}
\end{table*}

\subsection{The effect of length normalization}

It is generally agreed that applying length normalization before LDA improves its performance. In order to reexamine its positive effect we perform an experiment to compare the performance of length normalization followed by LDA and LDA without length normalization. The system with MFCC features and uncertainty normalization is used as the single system in this experiment. Table \ref{tbl.ln} shows that although cosine similarity scoring applies length normalization implicitly, applying length normalization before LDA and after uncertainty compensation is beneficial. As discussed above, length normalization makes vectors more normally distributed, which is in line with the Gaussian assumptions of LDA. 

\begin{table*}[th]
  \renewcommand{\arraystretch}{1.0}
  \caption{\label{tbl.ln}\it{Comparison between LDA with and without length normalization.}}
  \vspace{-2mm}
  \centerline
  {
    \begin{tabular}{l l  c c c c c c c c}
      \toprule
      & & & \multicolumn{3}{c}{\bfseries{Male}} & & \multicolumn{3}{c}{\bfseries{Female}} \\
      \cmidrule{4-6} \cmidrule{8-10}
      Method & & & EER [\%] & \oldmindcf & \newmindcf & & EER [\%] & \oldmindcf & \newmindcf \\
      \midrule
      \multirow{1}{*}{LDA with length normalization}
            	& & &  1.52 &  0.093 &  0.517			&  &  1.77 &  0.094 &  0.424 \\
      \midrule
      \multirow{1}{*}{LDA without length normalization}
                & & &  1.68 & 0.112 & 0.521			&  & 1.98 & 0.105 & 0.431 \\
      \bottomrule
    \end{tabular}
  }
  \vspace{-3mm}
\end{table*}

\subsection{Results on phrases using the RedDots corpus}
Although we developed our method primarily for words as recognition units, we can evaluate it on short phrases in a similar way. For experimentation on phrases, RSR2015 part I used to be a standard option, however it is now considered as a too easy corpus \cite{stafylakis2013text}. RedDots is more challenging in terms of channel variability, mostly due to (a) the longer time intervals between successive recordings of the same speaker, and (b) the higher levels of background noise \cite{lee2015reddots}. 

The main caveat of RedDots is the lack of a training set, due to the small number of participants (49 males and 13 females, with only 35 males and 6 females having target trials). This shortcoming prevents us from evaluating our method on the whole set of RedDots phrases, since training utterances of the evaluation phrases are compulsory in order to train our models. Nevertheless, two of the RedDots phrases (namely the 33rd and the 34th) are also contained in RSR2015 Part I, enabling us to train our models on the corresponding training utterances of RSR2015.

In Table \ref{tbl.red_dots_results} we report the performance on male-only trials, as the number of female speakers is too small for drawing any conclusion. The results are averaged over the 33rd and 34th RedDots phrases. Our focus is on the Impostor-Correct results, i.e. with non-targets trials containing the correct phrase, since we believe ASR-based methods are more adequate to estimate whether or not the uttered phrases match the prompted ones. The results show that uncertainty normalization is more effective than regularized LDA, attaining drastic relative improvement in low false acceptance operating points (63\% in \oldmindcf and 45\% in \newmindcf) and 30\% relative improvement in terms of EER. These improvements are attained without any channel compensation, i.e. without requiring repetitions of the same phrase from each training speaker. Finally, by combining uncertainty normalization with regularized LDA a further small improvement is attained.

\begin{table}[th]
  \renewcommand{\arraystretch}{0.9}
  \caption{\label{tbl.red_dots_results} The effectiveness of uncertainty normalization on male part of RedDots data (impostor-correct trials).}
  \vspace{-2mm}
  \centerline
  {
  	\setlength\tabcolsep{4pt}
    \begin{tabular}{c c c c c}
      \toprule
      Uncer. Norm. & Reg. LDA & EER [\%] & \oldmindcf & \newmindcf \\
      \midrule      
      	\xmark & \xmark & 3.12 & 0.239 & 0.381 \\
      \midrule      
      	\xmark & \cmark &	2.35 & 0.103 & 0.342 \\
      \midrule      
      	\cmark & \xmark &	2.19 & 0.088 & 0.209 \\
      \midrule      
      	\cmark & \cmark &	2.11 & 0.086 & 0.211 \\
      \bottomrule
    \end{tabular}
  	\vspace{-5mm}
  }
\end{table}

\subsection{Discussion}

The comparison of our results with the tied-mixture model approach (Table \ref{tbl.results}) shows that our single system outperforms the baseline by a large margin. In fact, its performance is superior not only to all single baseline systems, but also to the fusion of all systems in~\cite{stafylakis2015text}. Furthermore, the use of i-vectors rather than supervector size features ($\vec{z}$-vectors) makes our methods significantly faster. Additionally, memory requirements for each speaker are considerably lower than those of the baseline. Moreover, our system attains higher performance compared to the current state-of-the-art (which is based on DNNs \cite{zhong2017dnn}), even when a single system is used and without training on any external dataset (Fisher dataset is used to train the tandem feature system in~\cite{zhong2017dnn}). 

In terms of uncertainty and channel compensation methods, uncertainty normalization followed by length normalization and LDA is the more effective combination (Table \ref{tbl.bottleneck}). The results in Table \ref{tbl.results} show a consistency with respect to features (MFCC and PLP) and gender, while the experiments on RedDots reaffirm the effectiveness of the proposed sequence of transforms, yielding drastic improvements especially in the low false alarm area (Table \ref{tbl.red_dots_results}).  
In terms of front-end features, MFCC perform consistently better than PLP in both genders, while bottleneck features seem to be marginally effective, and only when fused with MFCC. Bottleneck features perform very well in text-independent speaker recognition, especially when used as a means to assign frames into UBM components \cite{lozano2016analysis} \cite{fu2014tandem}. However, there is a severe mismatch between the way frames are assigned in text-independent speaker recognition and our proposed HMM-based method. For example, the large context window used in the former does necessarily provide fine-grained temporal localization, required in order to segment each digit into $S_d = 8$ states. More recent end-to-end methods may be more effective ways of using DNNs for text-dependent speaker recognition than bottleneck features (e.g. \cite{zhang2016end}, \cite{chowdhury2017attention}), with the caveat that they require large amounts of in-domain data, which are not available in RSR2015 part III.

\subsection{Scaling-up to larger vocabulary}
In cases where a larger vocabulary can be employed the proposed method may suffer from data fragmentation. The number of overall training examples should scale linearly with the number of words, as no parameter sharing is assumed between the word-specific models. In such cases, introducing parameter sharing between the models (especially between the several word-specific HMMs and i-vector extractors) should be considered. Although such a setting is beyond the scope of this work, one may start with a typical large-size UBM/i-vector system (e.g. with 2048 Gaussian components) trained on text-independent datasets or on the available in-domain dataset. Then for each word in the vocabulary, the $C_d$ most dominant Gaussian components should be selected and their means should possibly be re-estimated e.g. via mean-only MAP adaptation, with the remaining components being removed. Word-specific i-vector extractors on top of the word-specific UBM can then be derived by (a) keeping only those rows corresponding to the $C_d$ most dominant Gaussian components, and (b) refining the matrix by applying e.g. minimum divergence training (i.e. without re-estimating the subspace). One may also consider starting from a higher dimensional i-vector extractor (e.g. 600) and selecting the most dominant dimensions for each word-specific extractor.

\section{Conclusions and Future Work}
\label{sec.conclusion}

In this paper, we developed a system for text-prompted speaker verification using random digit strings. The core of the system comprises a set of digit-specific HMMs, which we employ in order to perform segmentation of utterances into digits, alignment of frames to HMM states and extraction of Baum-Welch statistics. On top of these HMMs, digit-specific i-vector extractors are trained, enabling us to compare digit-specific i-vectors that appear in both enrolment and test utterances using simple cosine distance scoring with score normalization. Furthermore, we investigated three different methods for compensating channel and uncertainty and we concluded that the novel uncertainty normalization technique followed by LDA yields consistently superior performance. The proposed system outperforms the baseline by a large margin and yields superior performance compared to the current state-of-the-art, which is based on DNNs.

We also examined the use of bottleneck features and different types of cepstral features. The experiments showed that although the performance of cepstral features is superior to that of bottleneck features, fusion with other cepstral features leads to further notable improvement. Our final set of experiments were conducted on whole phrases. To this end, the challenging RedDots corpus is used \cite{lee2015reddots}. The results we reported reaffirm the effectiveness of uncertainty normalization, yielding an impressive 63\% relative improvement in terms of \oldmindcf.

For future work, we are interested in fitting certain elements of the proposed approach to end-to-end neural architectures. Recently emerged approaches in text-independent speaker recognition combine end-to-end deep learning methods with implicit modeling of acoustic units via multi-head attention and learnable dictionaries or with mimicking the i-vector/PLDA framework \cite{zhu2018self,cai2018exploring,rohdin2018end}. We expect that the proposed method will contribute to this research direction, by demonstrating the potential of digit-specific HMMs and i-vector extractors. 

Finally, we should note that the channel and uncertainty compensation approaches examined here may also be applicable to speaker embeddings. Modeling the uncertainty in x-vectors is less straightforward compared to i-vectors. However, recent advances in Bayesian deep learning demonstrate that model averaging via dropouts is a means for quantifying the uncertainty of extracted representations \cite{gal2016dropout}. As a result, uncertainty normalization may also be relevant to neural representations, such as x-vectors.

\section{Acknowledgements}
Themos Stafylakis is funded by the European Commission program Horizon 2020, under grant agreement no. 706668 (Talking Heads).


\bibliographystyle{IEEEtran}
\bibliography{mybib}

\vspace{-8mm}

\begin{IEEEbiography}[{\includegraphics[width=1in,height=1.25in,clip,keepaspectratio]{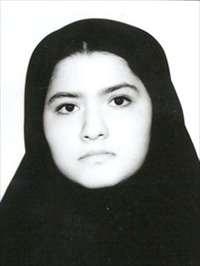}}]{Nooshin Maghsoodi}
Nooshin Maghsoodi received the B.Sc. degree in Computer Engineering from Sharif Universiy of Technology and M.Sc. in Artificial Intelligence from AmirKabir University of Technology. Since 2013, she is a Ph.D. student in Artificial Intelligence at Sharif University of Technology. Her interest are speaker verification, graphical models and machine learning.
\end{IEEEbiography}

\vspace{-10mm}
\begin{IEEEbiography}[{\includegraphics[width=1in,height=1.25in,clip,keepaspectratio]{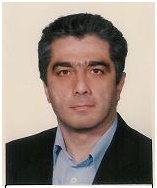}}]{Hossein Sameti}
Dr. Hossein Sameti received his Ph.D. degree in Electrical Engineering from University of Waterloo, Canada, in 1994. In 1995, he joined the Department of Computer Engineering, Sharif University of Technology, where he is an Associate Professor now. He founded Speech Processing Lab (SPL) at the department in 1998 and is the supervisor of the lab. Dr. Sameti's current research interests include speech  processing, automatic speech recognition, speech synthesis, speech enhancement, spoken dialogue systems, spoken language understanding, speaker verification, and spoken term detection.
\end{IEEEbiography}

\vspace{-8mm}
\begin{IEEEbiography}[{\includegraphics[width=1in,height=1.25in,clip,keepaspectratio]{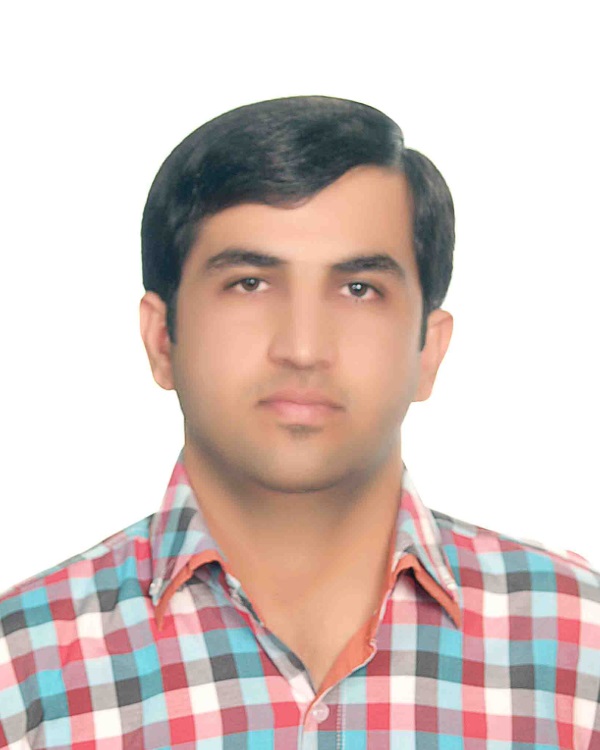}}]{Hossein Zeinali}
Dr. Hossein Zeinali received his B.Sc. degree in Computer Engineering from Shiraz University, Shiraz, Iran in 2010, his M.Sc. and Ph.D. degrees in Artificial Intelligence from Sharif University of Technology, Tehran, Iran in 2012 and 2017 respectively. He is currently a post-doc researcher at speech group of the Brno University of Technology. His research interests include speech processing, speaker recognition, deep learning, and machine learning.
\end{IEEEbiography}

\vspace{-10mm}
\begin{IEEEbiography}[{\includegraphics[width=1in,height=1.25in,clip,keepaspectratio]{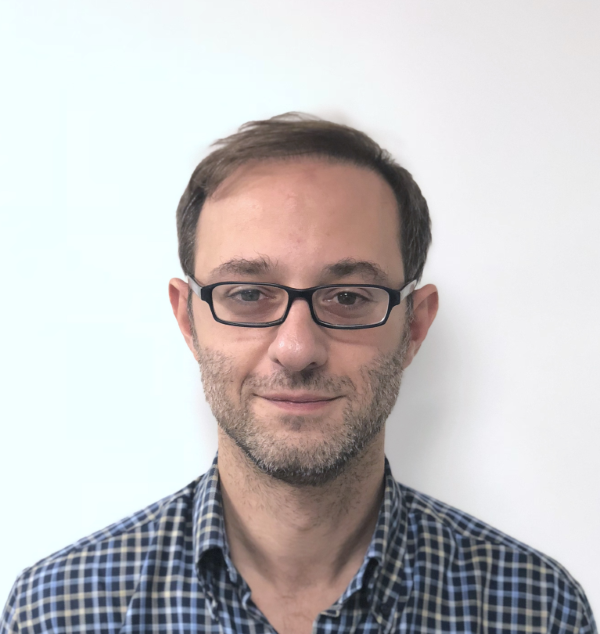}}]{Themos Stafylakis}
Dr. Themos Stafylakis received his PhD in Speaker Diarization for Broadcast News from National Technical University of Athens, Greece in 2011, his M.Sc. in Communication and Signal Processing from Imperial College London, UK in 2005 and his B.Eng. from National Technical University of Athens, Greece in 2004. In 2011 he joined CRIM (Canada) as a post-doc researcher on speaker recognition. In 2016 he joined the Computer Vision Laboratory at University of Nottingham (UK), as Marie Curie Research Fellow. His main research interests are audiovisual speech and speaker recognition, machine learning and deep neural networks.
\end{IEEEbiography}

\end{document}